\documentclass[11pt,a4paper,aps,nofootinbib]{article}
\usepackage{jcappub}
\usepackage[utf8]{inputenc}
\usepackage{color}
\usepackage{amsmath, bm, physics}
\usepackage{amssymb, mathtools}
\usepackage{subcaption}
\usepackage{hyperref}

\title{Strong scale-dependence does not enhance the kinematic boosting of gravitational wave backgrounds}

\author[a,1]{G. Mentasti,\note{Corresponding author.}}
\author[a]{C.~R. Contaldi,}
\author[b,c]{and M. Peloso}

\affiliation[a]{Blackett Laboratory, Imperial College London, SW7 2AZ, United Kingdom}
\affiliation[b]{Dipartimento di Fisica e Astronomia ``G. Galilei", Universit\`a degli Studi di Padova, via Marzolo 8, I-35131 Padova, Italy}
\affiliation[c]{INFN, Sezione di Padova, via Marzolo 8, I-35131 Padova, Italy}

\emailAdd{g.mentasti21@imperial.ac.uk}

\abstract{
Existing expressions in the literature appear to indicate that Doppler boosting, due to our proper motion with respect to the isotropic frame of the universe, can amplify stochastic gravitational wave backgrounds whose energy spectra exhibit strong scale dependence, for example, those generated by large scalar perturbations in models of primordial black holes or by astrophysical populations with broken power-law behaviour. It has been suggested that this enhancement could increase the signal-to-noise ratio of such backgrounds in pulsar timing measurements, as well as in ground- and space-based observatories.
We show that the reported enhancement is an artefact of a Taylor expansion of the boosted signal, typically performed in the literature under the assumption of a small boosting parameter. This approximation fails to reproduce the correct result for signals with strong scale dependence. When Doppler boosting is treated exactly, the apparent amplification disappears. Using representative spectra, we demonstrate that Doppler motion induces only blue- and red-shifting by the expected amount; it does not lead to additional amplification or introduce new spectral features. The exact expression for the kinematic boost can and should be easily applied in analysing such backgrounds. 
}

\begin{document}
	
\maketitle
	
\section{Introduction}\label{sec:introduction}

Cosmology relies on the statement that our Universe is homogeneous and isotropic at large scales, as indicated by data. The assumption of isotropy should be better qualified, namely, one should more properly state that there exists a reference frame in which the Universe appears statistically isotropic at large scales. Observers not at rest in that frame will find their measurements influenced by their motion relative to it. This has been experimentally demonstrated in Cosmological Microwave Background (CMB) measurements~\cite{Smoot:1977bs, Kogut:1993ag,WMAP:2003ivt, Planck:2013kqc}. To leading order, boosting the monopole signal of the isotropic frame results in a dipole component \cite{Challinor:2002zh}, from which we learn that the Solar System is moving in this frame with a velocity of magnitude $\beta \equiv v / c = \left( 1.23 \pm 0.003 \right) \times 10^{-3}$ in the direction of Galactic longitude and latitude $\left( \ell ,\, b \right) \simeq \left( 264^\circ ,\, 48^\circ \right)$. Besides the dipole, the boosting also generally results in the generation of higher multipole anisotropies from lower multipole ones, and in the modulation and aberration of the intensity anisotropies present in the statistically isotropic frame, effects that have also been measured in the CMB data~\cite{Planck:2013kqc}. 

The assumption of statistical isotropy has been questioned by claims of an existing tension between the CMB dipole and the dipole observed in source counts of quasars and radio galaxies~\cite{Secrest:2020has,Dalang:2021ruy,Secrest:2022uvx} (see, however, \cite{Darling:2022jxt}).
Data of a completely different nature might be beneficial in this discussion. In addition to the CMB, our Universe hosts at least two other stochastic backgrounds, namely those of cosmological neutrinos and gravitational waves.
While the detection of the former constitutes a formidable challenge~\cite{PTOLEMY:2019hkd}, enormous progress is currently being made in the field of observational gravitational waves, with evidence for a Stochastic Gravitational Wave Background (SGWB) emerging from the NANOGrav collaboration~\cite{NANOGrav:2023gor}. The origin of this background is still under debate.
More in general, a cosmological background of gravitational waves~\cite{Caprini:2018mtu} is expected to have a dominant isotropic component plus small ansisotropies originating from the propagation of the gravitational waves in the perturbed universe~\cite{Alba:2015cms,Contaldi:2016koz,Cusin:2017fwz,Jenkins:2018nty,Bartolo:2019oiq,Bartolo:2019yeu}, as well as from the intrinsic large scale anisotropies of their source, as explored in specific mechanisms~\cite{Geller:2018mwu,Bartolo:2019zvb}. Analogously to the CMB, it is reasonable to assume that, for an SGWB signal of cosmological origin, the kinematic dipole dominates over the one already present in the statistically isotropic frame. 

Assuming that the SGWB can be measured to sufficient accuracy and precision to reveal its kinematic dipole component, the comparison of the values of our peculiar velocity from CMB and SGWB data would result in a formidable test of the principle of the isotropy of the universe. Such a measurement is however not trivial, due to the ${\rm O } \left( \beta \right) = {\rm O } \left( 10^{-3} \right)$ suppression of an already small signal, and due to the intrinsic variation of the dominant monopole component that, even in the limit of an ideal (no noise) instrument, results in a non-vanishing value for all multipoles in any finite-duration experiment, even if the SGWB was perfectly statistically isotropic~\cite{Mentasti:2023icu}. For a scale-invariant signal, third-generation ground-based gravitational wave detectors will be able to measure the kinematic dipole only if the amplitude of the monopole component is very close to its current upper bound~\cite{Mentasti:2023gmg}. The situation is slightly better for LISA, although in this case the detection also appears to be particularly challenging due to the planar nature of the instrument, which suppresses its sensitivity to odd multipoles~\cite{LISACosmologyWorkingGroup:2022kbp}.

These challenges have led several authors to investigate whether better prospects for the detection of these kinematic effects can be found in the case of a strongly scale-dependent signal. The formula for the SGWB amplitude in the boosted frame presented in the literature~\cite{LISACosmologyWorkingGroup:2022kbp} appears to indicate this. Denoting by $\Omega_{\rm GW} \equiv \frac{1}{\rho_{\rm crit}} \, \frac{d \rho_{\rm GW}}{d \ln f}$ the fractional contribution per logarithmic frequency $f$ interval of the gravitational wave energy density, relative to that of a flat universe, and assuming that there exists a frame, indicated by prime, in which $\Omega_{\rm GW}'$ is isotropic, the measurement in a frame that moves with velocity $\vec{v} = \beta \, {\hat v}$ with respect to the isotropic frame gives
\begin{equation}
\Omega_{\rm GW} \left( f ,\, {\hat n} \right) = \Omega_{\rm GW}' \left( f \right) 
\left\{ \left[ 1 + {\cal M}_\Omega \left( f \right) \right] + {\hat n} \cdot {\hat v} \, {\cal D}_\Omega \left( f \right) + \left[ \left( {\hat n} \cdot {\hat v} \right)^2 - \frac{1}{3} \right] {\cal Q}_\Omega \left( f \right) + {\rm O } \left( \beta \right)^3 \right\} \;, 
\label{boost}
\end{equation} 
where ${\hat n}$ is the line-of-sight direction and where the modulating functions~\cite{LISACosmologyWorkingGroup:2022kbp} 
\begin{eqnarray}
{\cal M}_\Omega \left( f \right) &\equiv& \frac{\beta^2}{6} \left[ 8 + n_\Omega \left( n_\Omega - 6 \right) + \alpha_\omega \right] \;, \nonumber\\
{\cal D}_\Omega \left( f \right) &\equiv& \beta \left( 4 - n_\Omega \right) \;, \nonumber\\
{\cal Q}_\Omega \left( f \right) &\equiv& \beta^2 \left( 10 - \frac{9 \, n_\Omega}{2} + \frac{n_\Omega^2}{2} + \frac{\alpha_\Omega}{2} \right) \;, 
\label{MDQ-Omega}
\end{eqnarray} 
are sensitive to the spectral shape of the SGWB signal
\begin{equation}\label{eq:n_Omega_alpha_Omega}
n_\Omega \left( f \right) \equiv \frac{d \, \ln \, \Omega_{\rm GW}' \left( f \right)}{d \, \ln \, f} \;\;,\;\; 
\alpha_\Omega \left( f \right) \equiv \frac{d \, n_\Omega \left( f \right)}{d \, \ln \, f} \;\;, 
\end{equation} 
suggesting a possible enhancement in the case of a strongly scale-dependent spectrum~\cite{Cusin:2022cbb}. 

It is important to emphasise that the relation (\ref{boost}) is a perturbative expansion in $\beta \ll 1$, justified by the above-mentioned value $\beta = {\rm O } \left( 10^{-3} \right)$ of the peculiar velocity of the solar system. Due to the smallness of this quantity, Doppler and aberration effects are almost always treated perturbatively in the literature.
For example, a first-order expansion already captures the kinematic dipole of the CMB and
the $\ell\!\leftrightarrow\!\ell\pm1$ mode couplings it induces, and it is accurate at the
$10^{-6}$ level for the temperature power spectrum
\cite{Planck:2013kqc,Challinor2002,KosowskyKahniashvili2011,NotariQuartin2012}.
Carrying out the full Lorentz transformation would therefore add complexity with no practical gain at current sensitivities.

Large-scale structure analyses also adopt the same logic.
Redshift-space distortions are modelled with linear-order velocity expansions
\cite{Kaiser1987,Hamilton1998}.
Pulsar-timing arrays likewise implement only first-order barycentric
corrections in their timing models
\cite{HellingsDowns1983,EdwardsHobbsManchester2006}.

Following this treatment, recent studies of the SGWB have expanded the Doppler-boosted
spectrum in a Taylor series in $\beta$
\cite{LISACosmologyWorkingGroup:2022kbp,Cusin:2022cbb,ChowdhuryTasinatoZavala2023,ChungJenkinsRomanoSakellariadou2022}.
The approximation is analytically convenient, maintaining a simple dependence on $\beta$, which aids parametric constraints, and, when considering anisotropies, maintains a simple coupling structure between scales \cite{Challinor:2002zh}, making spherical harmonic treatments tractable. However, here we show that it converges only when the spectrum is sufficiently smooth,
i.e. when the logarithmic derivatives of the spectrum are finite over the band of interest. The CMB case, which is often used as an analogy to the SGWB, is a particular case where a thermal spectrum is assumed when considering the kinematic effects. In this case, the expansion in small $\beta$ is justified, leading to well-understood contributions to the anisotropies. In contrast, SGWBs can be sourced by several mechanisms, which can result in 
broken power-law spectra or spectra with sharp features that violate this bound,
causing the truncated series to mimic a spurious enhancement of the boosted signal, which disappears in the exact Lorentz calculation. This observation is the central result of this work.~\footnote{As it will be clear from our computations, this statement applies not only to the SGWB but to any cosmological relativistic background.} 

This {\sl paper} is structured as follows. In Section~\ref{sec:Omega_boosted} we provide a full derivation of the boosted signal. Although the derivation is analogous to that of ~\cite{LISACosmologyWorkingGroup:2022kbp}, and the results agree, we perform computations in terms of the phase-space energy density of the SGWB, which has a more immediate interpretation than the quantity $\Omega$ used in that work. After obtaining the exact expression for the boosted signal, in Subsection~\ref{sec:Taylor} we Taylor expand it in small $\beta$.
In Section~\ref{sec:comparison}, we compare with the existing literature by first showing that our derivation reproduces the result~\eqref{MDQ-Omega} of~\cite{LISACosmologyWorkingGroup:2022kbp}, and by then listing a series of works that used or discussed these expressions in the case of strongly scale-dependent signals.
In Section~\ref{sec:Taylor-limitation}, we first add the next ${\rm O } \left( \beta^3 \right)$ term to the expression of the boosted dipole. The addition shows the core of the problem: higher-order terms in $\beta$ are multiplied by higher derivatives of the spectral function, invalidating the accuracy of the $\beta-$truncated series for strongly scale-dependent signals.
We then present two subsections in which we compare the exact and Taylor-expanded boosted signal (i) for a simple toy spectrum, where also the exact expression has a simple analytical form, and (ii) for the case of a SGWB induced at second order by enhanced scalar perturbations, as this mechanism has been used as the main example of this enhancement in the literature. Both examples explicitly show that the enhancement is an artefact of the Taylor expansion, and that the exact boosted expression retains an amplitude comparable to that of the signal in the original frame (as one would have expected, given that we work in the limit of very small boost parameter, as indicated by the CMB dipole). Finally, Section~\ref{sec:conclusions} contains our conclusions. 

\section{A derivation of the boosted anisotropies}\label{sec:Omega_boosted}

We denote by $\mathcal{S}'$ the frame in which the SGWB is statistically isotropic. In this frame, the quantities $\hat n'$ and $f'$ refer, respectively, to the sky directions and frequencies of the measured signal. We denote by $\mathcal{S}$ the frame of the observer, and by $\hat n$ and $f$ the sky directions and frequencies in this frame. Both frames are assumed to be inertial, and, as mentioned in the Introduction, we denote by $\vec{v} = \beta \, {\hat v}$ the velocity of $\mathcal{S}$ with respect to $\mathcal{S}'$. The Doppler shift induced by this relative motion is quantified by the direction-dependent operator
\begin{align}
{\mathcal D}(\hat n')&\equiv\frac{1+\beta\,\hat n'\cdot\hat v}{\sqrt{1-\beta^2}}\,.
\end{align}
The frequencies in the two frames are related by the transformation
\begin{align}
f&=\mathcal{D}(\hat n')f'\,,
\end{align}
which is inverted by the substitution $\hat v\to -\hat v$, resulting in 
\begin{align}
f'&= \mathcal{D}(-\hat n)f\,.
\label{fpf}
\end{align}
In the statistically isotropic frame, the number of gravitons $d N'$ in the infinitesimal phase-space element defined by the range of frequency $(f',f'+\mathrm{d}f')$, the line of sight interval $(\hat n',\hat n'+\mathrm{d}^{2}\hat n')$ and the physical volume $\mathrm{d}V'$ is 
\begin{align}
dN'&=\Delta'(f')\,f'^2\,df'\,d^2\hat n'\,dV'\,,
\end{align}
with $\Delta'(f')$ defining the spectral number density of gravitons, which is assumed to be homogeneous and isotropic in the $\mathcal{S}'$ frame, where therefore it does not depend on $\hat n'$, nor on the spatial coordinates. Analogously, the boosted observer in $\mathcal{S}$ observes
\begin{align}
dN&=\Delta(f,\hat n)\,f^2\,df\,d^2\hat n\,dV\,.
\end{align}
The particle number is Lorentz invariant (\(\mathrm{d}N' = \mathrm{d}N\)).
Using the relations between solid angles and volumes under a Lorentz boost,~\footnote{See~\cite{1979AmJPh..47..602M} for details.}
\begin{align}
d^2\hat n& = d^2\hat n' / \mathcal{D}^2 \left( \hat n' \right) \,,\nonumber\\
dV& = dV' / \mathcal{D} \left( \hat n' \right) \,,
\end{align}
we obtain
\begin{align}
\Delta'(f')&=\Delta(f,\hat n)\,.
\end{align}
To derive the spectral properties of the SGWB from first principles, we consider the energy budget of the gravitons, per volume and element of the momentum space. See Section~\ref{sec:comparison} for details. In the reference frame $\mathcal{S}'$ (an analogous formula holds for $\mathcal{S}$) we have
\begin{align}
dE'=f'\,dN'=\Delta'(f')\,f'^3\,df'\,d^2\hat n'\,dV'\,,
\end{align}
and the energy density
\begin{align}\label{drho}
d\rho'=\frac{dE'}{dV'}=\Delta'(f')\,f'^3\,df'\,d^2\hat n'\,.
\end{align}
We introduce the SGWB in the $\mathcal{S}$ reference system (the definition in the $\mathcal{S}'$ system is analogous)
\begin{align}\label{hbackground}
h_{ab}(t,\vec x)=\sum_\lambda\int_{-\infty}^{+\infty} df\int d^2\hat n \, h_\lambda \left( f,\hat n \right) \, e^{2\pi i f \left( t-\hat n\cdot\vec x/c \right) } \, e^\lambda_{ab} \left( \hat n \right) \,,
\end{align}
where $\lambda$ runs over all allowed polarisations defined by the tensors $e_{ab}^\lambda \left( \hat n \right)$, normalized according to $e^{\lambda*}_{ab}(\hat n)\,e^{\lambda'}_{ab}(\hat n') = \delta_{\lambda\lambda'}$. Assuming a homogeneous, stationary, and unpolarized statistics for the SGWB
\begin{align}
\left\langle h_\lambda^{\,}(f,\hat n)h^\star_{\lambda'}(g,\hat m) \right\rangle = \delta_{\lambda\lambda'} \, \delta(f-g) \, \frac{\delta^{(2)} \left( {\hat n} - {\hat m} \right)}{4 \pi} \, \mathcal{H}(f,\hat n)\,.
\label{hh}
\end{align}
For simplicity, we also assume that the power spectral dependence and anisotropy $\mathcal{H}(f,\hat n)$ can be factorised as
\begin{align}
\mathcal{H}(f,\hat n)=H(f) \, P(\hat n)\,.
\end{align}
where we normalise $P(\hat n) = 1$ in the isotropic case.~\footnote{This choice, as well as the normalisation of the polarisation operator and the $4 \pi$ factor in Eq.~\eqref{hh}, follow the conventions of~\cite{LISACosmologyWorkingGroup:2022kbp}.}

It can be shown that the total energy density in the SGWB is 
\begin{align}\label{rhotot}
\rho=\frac{c^2}{8 \, G} \int_0^{+\infty} df \int d^2\hat n\,f^2\,\mathcal{H}(f,\hat n)\,,
\end{align}
where $G$ is Newton's constant. Therefore, using eqs.~\eqref{drho} and~\eqref{rhotot}, we can identify
\begin{align}
\mathcal{H}(f,\hat n)=\frac{8\,G}{c^2}f\,\Delta(f,\hat n)\,.
\end{align}
This determines how the spectral power density transforms from the isotropic reference frame $\mathcal{S}'$, where $\mathcal{H}'(f',\hat n')=H'(f')$, to the boosted frame $\mathcal{S}$ with redshifted frequencies and induced anisotropies
\begin{align}\label{Hcal_transform}
\mathcal{H}^{\rm exact}(f,\hat n)=\mathcal{D}(\hat n')\,H'(f')&= \frac{H'(\mathcal{D} \left( -\hat n \right) \, f)}{\mathcal{D} \left( -\hat n \right)} \,.
\end{align}

\subsection{Taylor expansion of the GW spectrum} \label{sec:Taylor}

In analogy to the approach taken in~\cite{LISACosmologyWorkingGroup:2022kbp}, we expand the exact transformation law \eqref{Hcal_transform} up to second order in $\beta \ll 1$:
\begin{align}\label{H_approx}
\mathcal{H}^{\rm approx}(f,\hat n)&\simeq H'(f)+\beta\left[u H'(f)- u f \frac{dH'(f)}{df}\right]\nonumber\\
&+\beta ^2 \left[\frac{1}{2} \left(f^2 u^2 \frac{d^2H'(f)}{d f^2}+f \frac{dH'(f)}{df}\right)-u^2 f \frac{dH'(f)}{d f}+\left(u^2-\frac{1}{2}\right) H'(f)\right] + {\rm O } \left( \beta^3 \right) \,,
\end{align}
where $u=\hat n\cdot\hat v$. We can rewrite this as
\begin{align}
\mathcal{H}^{\rm approx}(f,\hat n)&\simeq H'(f) \Bigg\{ 1 + \beta \, u \left[ 1 - n_{\cal H}\right] \nonumber\\
&+\beta ^2 \left[ \frac{u^2}{2} \, \alpha_{\cal H} + \frac{1}{2} \left[ 1 - 3 u^2 + u^2 n_{\cal H} \right] n_{\cal H} + \left( u^2 - \frac{1}{2} \right) \right] + {\rm O } \left( \beta^3 \right) \Bigg\} \,,
\end{align}
where the spectral indices are defined as
\begin{equation}
n_{\cal H} \left( f \right) \equiv \frac{d\ln H' \left( f \right)}{d\ln f} \;\;,\;\; 
\alpha_{\cal H} \left( f \right) \equiv \frac{d\,n_{\cal H} \left( f \right)}{d\ln f}\,.
\label{eq:n_H_alpha_H}
\end{equation}

It is straightforward to rearrange the expression as monopolar, dipolar, and quadrupolar contributions to the induced anisotropy,
\begin{align}\label{eq:H_anis}
\mathcal{H}^{\rm approx}(f,\hat n)&=H'(f) \left[ 1+\mathcal{M}_{\cal H}(f)+u\,\mathcal{D}_{\cal H}(f)+\left(u^2-\frac 1 3\right)\mathcal{Q}_{\cal H}(f) + {\rm O } \left( \beta^3 \right) \right]\,,
\end{align}
where
\begin{align}
\mathcal{M}_{\cal H}(f)&\equiv\frac{\beta^2}{6}\left[n_{\cal H}^2(f)+\alpha_{\cal H}(f)-1\right]\,,\nonumber\\
\mathcal{D}_{\cal H}(f)&\equiv\beta\left[1-n_{\cal H}(f)\right]\,,\nonumber\\
\mathcal{Q}_{\cal H}(f)&\equiv\beta^2\left[\frac{\alpha_{\cal H}(f)}{2}-\frac{3n_{\cal H}(f)}{2}+\frac{n_{\cal H}^2(f)}{2}+1\right]\,.
\label{MDQ-H}
\end{align}

\section{Comparison with the previous literature} \label{sec:comparison}

Here we study the (exact, and Taylor-expanded) expressions for the transformation under a boost of the quantity ${\cal H}$ defined through Eq.~\eqref{hh}. As manifest from Eq.~\eqref{rhotot}, this quantity (up to the $c^2/ 8 G$ rescaling) is the integrand of the SGWB energy density in momentum space. In~\cite{LISACosmologyWorkingGroup:2022kbp} the authors performed an analogous computation for the quantity $\Omega(f,\hat n)$, defined in \cite{LISACosmologyWorkingGroup:2022kbp} (by combining their eqs.~(C.5) and~(C.6)) through its relation to the gravitational wave energy density
\begin{align}
\rho = \rho_c \, \int d \ln f \int d^2 {\hat n} \, \Omega \left( f ,\, {\hat n} \right) \;,
\end{align}
where $\rho_c$ is the critical energy density required for a flat universe having the present Hubble rate $H_0$. 
Comparing this with Eq.~\eqref{rhotot}, we find
\begin{align}
\Omega \left( f ,\, {\hat n} \right) = \frac{c^2}{8\,G\rho_c} \, f^3\,\mathcal{H} \left( f,\hat n \right) = \frac{\pi}{3 H_0^2} \, f^3\,\mathcal{H} \left( f,\hat n \right)  \;,
\label{Omega-to-H}
\end{align}
where $H_0$ is the Hubble rate and the numerical prefactor follows from the normalisation conventions adopted in \cite{Allen:1996gp,Mentasti:2023gmg} and followed here.\footnote{See also, e.g. \cite{Moore:2014lga}. We introduce all anisotropic, spectral quantities as densities in both frequency and angular domains for consistency.} 
Namely, the combination $\Omega \left( f ,\, {\hat n} \right) / f^3$ is also the phase-space energy density up to an overall constant. Therefore, in the isotropic frame, $\ln \Omega' \left( f' \right) = \ln \mathcal{H}' \left( f' \right) + 3 \, \ln f' + {\rm constant}$, so that, the parameters in eqs.~\eqref{eq:n_Omega_alpha_Omega} and~\eqref{eq:n_H_alpha_H} are related to each other by 
\begin{equation}
n_{\cal H} \left( f \right) = n_\Omega \left( f \right) - 3 \;\;,\;\; \alpha_{\cal H} \left( f \right) = \alpha_\Omega \left( f \right) \;. 
\label{shift}
\end{equation}
We performed computations in terms of $\mathcal{H}$ rather than of $\Omega$, as the former quantity has a more immediate physical interpretation (energy density in phase space, rather than energy density times the frequency cubed). Nonetheless, our derivation mirrors that of~\cite{LISACosmologyWorkingGroup:2022kbp} and our result~\eqref{eq:H_anis} can be readily employed to reproduce Eq.~\eqref{boost} of ~\cite{LISACosmologyWorkingGroup:2022kbp}. To see, this, we combine eqs.~\eqref{fpf} and~(\ref{Hcal_transform}) to write 
\begin{align}
\Omega^{\rm approx} \left( f ,\, {\hat n} \right) &= \frac{8 \pi^2}{3 H_0^2} \, f^3 \, \left[ \frac{H' \left( {\cal D} \left( - {\hat n} \right) f \right)}{{\cal D} \left( - {\hat n} \right)} \right]_{\rm expanded} \,,\nonumber\\
&= \frac{8 \pi^2}{3 H_0^2} \, f^3 H'(f) \left[ 1+\mathcal{M}_{\cal H}(f)+u\,\mathcal{D}_{\cal H}(f)+\left(u^2-\frac 1 3\right)\mathcal{Q}_{\cal H}(f) + {\rm O } \left( \beta^3 \right) \right]\,,\nonumber\\
&= \Omega' \left( f \right) \left[ 1+\mathcal{M}_{\cal H}(f)+u\,\mathcal{D}_{\cal H}(f)+\left(u^2-\frac 1 3\right)\mathcal{Q}_{\cal H}(f) + {\rm O } \left( \beta^3 \right) \right]\,,\nonumber\\
&= \Omega' \left( f \right) \left[ 1+\mathcal{M}_\Omega (f)+u\,\mathcal{D}_\omega(f)+\left(u^2-\frac 1 3\right)\mathcal{Q}_\Omega (f) + {\rm O } \left( \beta^3 \right) \right]\,,  
\end{align}
where the last equality indeed reproduces Eq.~\eqref{boost} and it follows from performing the shift~\eqref{shift} in the expressions~\eqref{MDQ-H}, which are then seen to be identical to the expressions~\eqref{MDQ-Omega}. We note that the equality $\Omega^{\rm approx} \left( f,\, {\hat n} \right) /  \Omega' \left( f \right) = \mathcal{H}^{\rm approx}(f,\hat n) / H'(f)$ is not in contradiction with the fact that these two quantities are proportional to a non-trivial function of the frequency (and so they transform differently under a boost), because the expressions in the denominators, are not the (rescaled) densities measured in the ${\cal S}'$ frame, but the formal expressions for these quantities in the ${\cal S}'$ frame, expressed as a function of the frequency in the ${\cal S}$ frame. 

After the expressions~\eqref{boost} were derived in~\cite{LISACosmologyWorkingGroup:2022kbp}, the authors of~\cite{Cusin:2022cbb} suggested that, since the functions ${\cal M}_\Omega$, ${\cal D}_\Omega$, and ${\cal Q}_\Omega$ are proportional to the tilts $n_\Omega$ and $\alpha_\Omega$, the measured signal in the boosted frame could be significantly enhanced in the case of a strongly scale-dependent spectrum. Strong scale dependence is, for instance, encountered in the well-studied case of an SGWB induced at second order by large scalar perturbations, see e.g.~\cite{Domenech:2021ztg,LISACosmologyWorkingGroup:2025vdz} for reviews. This potential enhancement was further studied or discussed in several works~\cite{Dimastrogiovanni:2022eir,Chung:2022xhv,Chowdhury:2022pnv,Tasinato:2023zcg,Heisenberg:2024var,Cruz:2024svc,Cusin:2025xle}.

While we confirm the validity of the approximation~\eqref{boost} derived in~\cite{LISACosmologyWorkingGroup:2022kbp}, and agree that the statement about increased scale-dependence of kinematic effects is, in itself, correct, the central question is how large this increase can be, given the perturbative nature of these relations. In the next section, we show that a significant enhancement is simply an indication that the perturbative relations have broken down, rather than a physical effect that would substantially aid in detecting the kinematic contributions.

\section{Limitations of the Taylor expansion approach} \label{sec:Taylor-limitation}

The key message of this work is that the Taylor-series treatment introduced in \eqref{H_approx} and the resulting expressions such as \eqref{eq:H_anis} are only reliable when the expansion converges. We show in the following that this criterion fails exactly in scenarios considered by earlier studies, namely those in which steep gradients in the spectral density push the series beyond its domain of validity.
To support this claim with just one explicit computation, let us include one additional order in $\beta$ in the dipole of Eq.~\eqref{eq:H_anis}, to find
\begin{align}
\mathcal{D}_{\cal H} \left( f \right) = 
\beta \left[ 1 - n_{\cal H} \right] +& 
\frac{\beta^3}{10} \Bigg[ 1 - n_{\cal H} \left( 1 - n_{\cal H} + n_{\cal H}^2 \right)  + \alpha_{\cal H} \left( 1 - 3 n_{\cal H}\right) - \frac{d \alpha_{\cal H}}{d \ln f } \Bigg] + {\rm O } \left( \beta^5 \right) \;.
\end{align} 
Studies in the literature have so far considered only the ${\rm O} \left( \beta \right)$ term. This is justified provided that the additional ${\rm O} \left( \beta^3 \right)$ contribution that we have computed, as well as all the other terms, remain small. We thus see explicitly that the ``expansion in small $\beta$'' considered in the literature does not just depend on $\beta$ bring small, but on the product between $\beta$ and progressively higher derivatives $\frac{d^n \ln H' \left( f \right)}{\left( d \ln f \right)^n}$ remaining small. Signals with a strong scale-dependence, which have been argued to have a large kinematic enhancement based on the expressions~\eqref{MDQ-Omega}, precisely fail to remain within this regime of validity. 

In the following, we show the failure of convergence in two examples. The first is an SGWB with a toy spectral dependence, for which the exact signal in the ${\cal S}$ frame acquires a simple analytic expression. The second is the example often employed in these studies, namely a SGWB induced at second order by scalar perturbations, for which the exact kinematic dipole can be straightforwardly computed numerically.  

\subsection{Application: A toy model}

To warm up with a system that is fully analytically tractable and provides a simple result, we compare the exact Doppler-boosted spectrum with the spectrum obtained from the Taylor series approximation for a toy model. The approximated result is enhanced to infinite amplitude at a frequency where the original signal has the greatest spectral dependence. However, the comparison with the exact result shows that this enhancement is an artefact of the ``truncation in small $\beta$'' of the approximate treatment. 

Consider an SGWB signal that, in the isotropic frame ${\cal S}'$, has the following spectral dependence
\begin{align}\label{eq:sqrt_H}
H' \left( f' \right) = A_0 \, \left(1-\frac{f'}{f_0} \right)^{1/2}\,,
\end{align}
where $f_0$ and $A_0$ are an arbitrary pivot scale and an arbitrary amplitude, respectively.  
We can compare the exact induced, anisotropic spectrum of Eq.~\eqref{Hcal_transform} to the first-order contribution in $\beta$: 
\begin{align}
\mathcal{H}^{\rm exact} \left(f , {\hat n} \right) &=A_0 \, \frac{\sqrt{1-\beta ^2}}{1-\beta \, u} \left[ 1 -\frac{1- \beta \, u}{\sqrt{1-\beta ^2}} \frac{f}{f_0} \right]^{1/2}\,,
\label{eq:exactsqrt}\\
\mathcal{H}^{\rm approx} \left(f , {\hat n} \right)  &= H' \left( f \right) \left[ 1 + \frac{u \, \beta}{2} \, \frac{2 - f/f_0}{1-f/f_0} + {\rm O } \left( \beta^2 \right) \right] \,. 
\label{eq:approxsqrt}
\end{align}
We note that the exact result~\eqref{eq:exactsqrt} is well behaved at $f\simeq f_0$, while the approximation~\eqref{eq:approxsqrt} diverges for any finite $\beta$. In fact
\begin{equation}
\lim_{f \to f_0 } \frac{d^n \ln H' \left( f \right)}{\left( d \ln f \right)^n} = - \frac{\left( n - 1 \right)!}{2} \, \left( \frac{1}{1-f/f_0} \right)^n \,,\;\;\; \mbox{for}\;\; n \geq 1 \;\;, 
\end{equation}
namely, progressively higher derivatives are progressively more divergent as $f$ approaches $f_0$, regardless of the values of $\beta$ or $u$, making the Taylor approximation invalid and making it appear to produce an enhanced amplitude.

\begin{figure}[th!]
\centering
\includegraphics[width=0.7\textwidth]{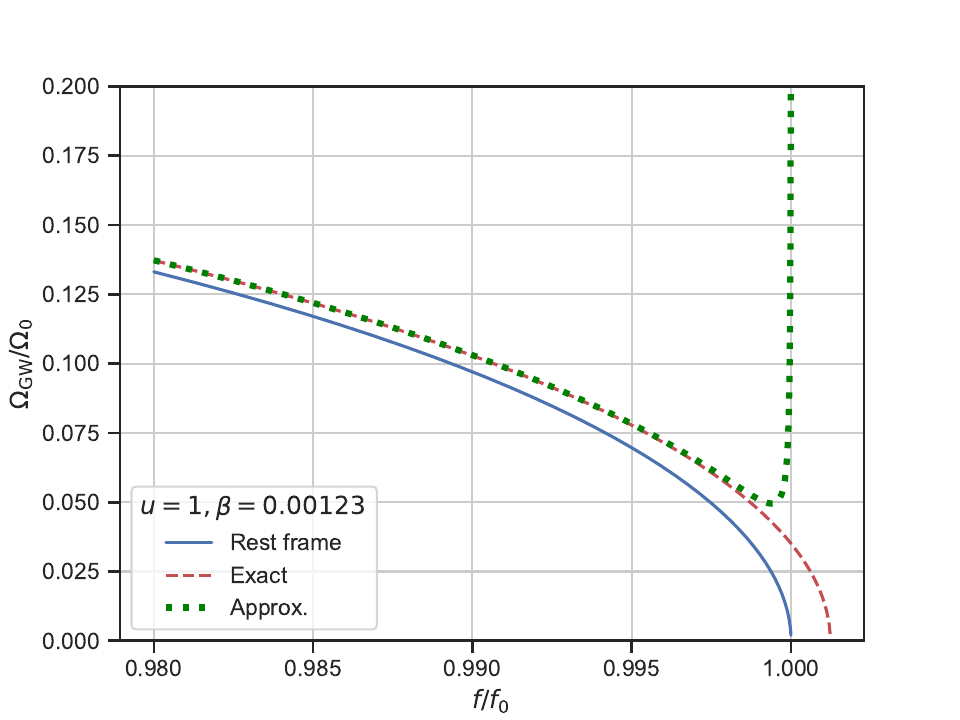}
    \caption{Toy model GW energy density from the power spectrum of Eq.~\eqref{eq:sqrt_H} in the isotropic frame of the sources (written in terms of $f$ - solid blue curve), the observed spectrum in the direction $u=\hat v\cdot\hat n=1$ (doshed orange) and its first order Taylor expansion in the kinematic velocity $\beta$ (dotted green). We take the boost factor $\beta =0.00123$, corresponding to the Solar System speed in the CMB rest frame.}
    \label{fig:sqrt}
\end{figure}

In Figure \ref{fig:sqrt} we show the spectrum of the signal in the isotropic ${\cal S}'$ frame (written in terms of $f$) and in the boosted frame ${\cal S}$, for the specific direction choice $u \equiv {\hat v} \cdot {\hat n} = 1$. Concerning the latter, we show both the exact~\eqref{eq:exactsqrt} and approximate~\eqref{eq:approxsqrt} expressions. To be more precise, we plot the function $\Omega_{\rm GW}(f,\hat n)\propto f^3 \, \mathcal{H}(f,\hat n)$ (see Eq.~\eqref{Omega-to-H}), in units of $\Omega_0 \equiv \pi A_0 f_0^3/3 H_0^2$. 

\subsection{Application: Scalar induced GWs}

\begin{figure}[h!]
    \centering
    \makebox[\linewidth][c]{%
        \begin{subfigure}[b]{0.5\textwidth}
            \centering
            \includegraphics[trim=0cm 0cm 0.2cm 0cm,clip=true,width=\textwidth]{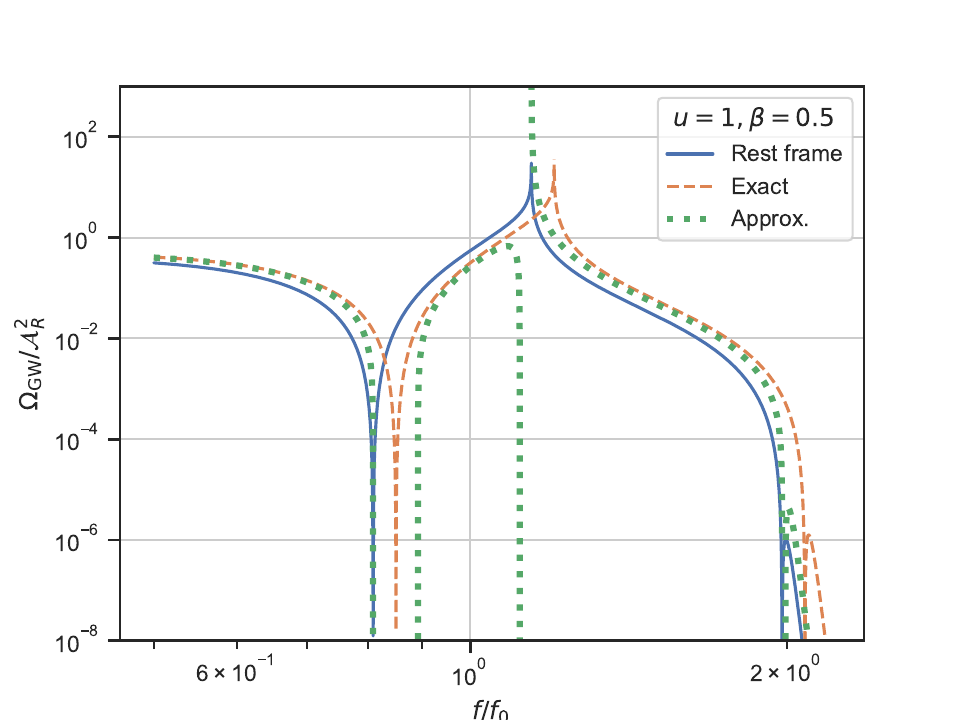}
        \end{subfigure}
        \hspace{0.1em}
        \begin{subfigure}[b]{0.5\textwidth}
            \centering
            \includegraphics[trim=0.5cm 0cm 0cm 0cm,clip=true,width=\textwidth]{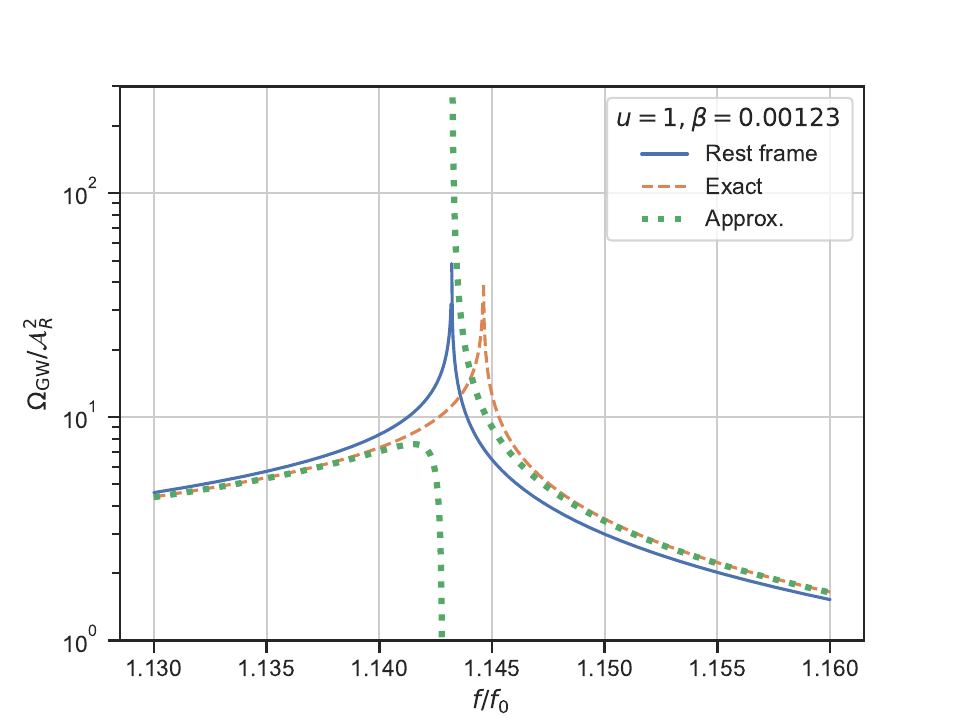}
        \end{subfigure}
    }
    \caption{Scalar induced power spectrum of Eq. \eqref{eq:PBH_Omega} in the reference frame of the sources (solid blue), the observed spectrum in the direction $u \equiv \hat v\cdot\hat n=1$ (dashed orange) and its first order Taylor expansion in the kinematic velocity $\beta$ (dotted green). A log-normal mass function with parameter $\Delta=0.1$ is considered. The plot of the left panel assumes a large value for the kinematic velocity $\beta=0.5$ to emphasise the discrepancy between the curves, while the right-hand panel shows a detail of the curves around the cusp of the GW spectrum with kinematic boost factor $\beta=0.00123$ given by the velocity of the Solar System in the CMB reference frame.}
    \label{fig:PBH}
\end{figure}

Primordial scalar curvature perturbations, responsible for the observed CMB anisotropies and large-scale structure formation, can also generate an SGWB through the standard non-linear gravitational interactions in the early universe~ \cite{Tomita:1975kj,Matarrese:1992rp,Matarrese:1993zf,Matarrese:1997ay,Acquaviva:2002ud,Mollerach:2003nq,Carbone:2004iv,Ananda:2006af,Baumann:2007zm,Domenech:2021ztg,LISACosmologyWorkingGroup:2025vdz}. While the amplitude of these induced GWs is negligible on large scales due to the smallness of the amplitude of the scalar perturbations $(\sim 10^{-5})$, constraints on much smaller scales are weaker, allowing for the possibility of enhanced scalar power that can induce an observable SGWB.~\footnote{These enhanced scalar perturbations can also lead to primordial black hole (PBH) formation. In this case, the viable asteroid-mass window for PBH dark matter is associated with an induced stochastic gravitational wave background (SGWB) that will be observed by LISA.~\cite{Bartolo:2018evs}.} For a peaked (as a function of wavenumber) spectrum of the sourcing scalar perturbations, the SGWB exhibits a strong scale-dependence in some frequency range. For this reason, such a mechanism has often been employed to exemplify the claimed SGWB enhancement from kinematic effects. The authors in~\cite{Pi:2020otn} considered a log-normal peaked spectrum of the curvature perturbations around a frequency $f_0$
\begin{align}
\mathcal{P}_{\mathcal{R}}(\kappa)=\frac{\mathcal{A}_{\mathcal{R}}}{\sqrt{2 \pi} \Delta} \exp \left(-\frac{\ln ^2\kappa}{2 \Delta^2}\right)\,,
\end{align}
where $\Delta$ is the dimensionless width of the spectrum, $\mathcal{A}_{\mathcal{R}}$ its normalization, and $\kappa \equiv \frac{f}{f_0}$. At small widths $\Delta \ll 1$, the induced SGWB spectrum admits the following analytic expression~\cite{Pi:2020otn}
\begin{align}\label{eq:PBH_Omega}
\!\!\!\! \!\!\!\! \!\!\!\! 
&\Omega_{\rm GW}(\kappa) \approx  3 \mathcal{A}_R^2 \kappa^2 e^{\Delta^2}\left[\operatorname{erf}\left(\frac{1}{\Delta} \text{sinh}^{-1} \frac{\kappa e^{\Delta^2}}{2}\right)-\text{erf}\left(\frac{1}{\Delta} \text{Re}\left(\text{cosh}^{-1} \frac{\kappa e^{\Delta^2}}{2}\right)\right)\right]\times\nonumber\\
&\left(1-\frac{\kappa^2 e^{2 \Delta^2}}{4} \right)^2\left(1-\frac{3 \kappa^2 e^{2 \Delta^2}}{2}\right)^2\times\\
&\left\{\left[\frac{1}{2}\left(1-\frac{3}{2} \kappa^2 e^{2 \Delta^2}\right) \ln \left|1-\frac{4}{3 \kappa^2 e^{2 \Delta^2}}\right|-1\right]^2+\frac{\pi^2}{4}\left(1-\frac{3}{2} \kappa^2 e^{2 \Delta^2}\right)^2 \Theta\left(2-\sqrt{3} \kappa e^{\Delta^2}\right)\right\}\,.\nonumber
\end{align}
Starting from this analytic expression, assumed to hold in the isotropic ${\cal S}'$ frame, we expect to observe the boosted signal shown in Figure~\ref{fig:PBH} in the boosted frame ${\cal S}$. Once again, we compare the exact expression to the approximation given in Eq.~\eqref{boost} and find that the enhancement is an artefact of the truncated series\footnote{The authors in~\cite{Pi:2020otn} explicitly show an extremely good agreement between the exact induced SGWB (obtained from a numerical integration) and the analytic approximation~\eqref{eq:PBH_Omega} for $\Delta = 10^{-2}$. A worse agreement is expected for the $\Delta = 10^{-1}$ choice that we employ in Figure~\ref{fig:PBH}. This issue is irrelevant for our study, where we focus on the comparison of the exact and approximated computation of the kinematic effect, in which Eq.~\eqref{eq:PBH_Omega} is just treated as an input function.}. Also in this case, the exact result is not enhanced with respect to the one shown in~\eqref{eq:PBH_Omega}. 

\section{Conclusions} \label{sec:conclusions}

Recent advances in observational gravitational wave astronomy have opened a new window of increasing sensitivity onto astrophysics, cosmology, and fundamental physics. Pulsar timing measurements have provided evidence for an SGWB at nanohertz frequencies~\cite{NANOGrav:2023gor}. Ground-based detectors are steadily improving sensitivity in the $\sim 10$–$100$ Hz range~\cite{KAGRA:2021kbb}, with a detection anticipated in the near future. The space-born mission LISA~\cite{LISA:2024hlh}, scheduled for launch in the next decade, will probe the millihertz regime.

Given the observed statistical isotropy of our universe in the CMB rest frame, it is natural to expect that the stochastic gravitational wave background (SGWB) is also isotropic in that frame. Experimentally verifying this expectation would provide a powerful test of the isotropy principle. However, achieving the accuracy necessary to detect this kinematic effect is challenging, due to the ${\rm O } \left( \beta \right) = {\rm O } \left( 10^{-3} \right)$ suppression of a monopole signal that is itself difficult to observe. For a scale-invariant SGWB, third-generation ground-based detectors will be able to detect the kinematic dipole only if the amplitude of the monopole component lies near its current upper bound~\cite{Mentasti:2023gmg}.

When performing these studies, it is important to recognise that the amplitude of the boosted signal depends on the frequency profile—or “spectral shape”—of the SGWB in the isotropic frame. Unlike the CMB, which follows a black-body spectrum, the spectral shape of the SGWB is model-dependent, and it is therefore still unknown. The relations derived in~\cite{LISACosmologyWorkingGroup:2022kbp}, shown in eqs.~\eqref{boost} and~\eqref{MDQ-Omega}, suggest that the boosted signal can be enhanced in the presence of a strongly scale-dependent SGWB. This has led to the speculation that 
a SGWB with strong scale dependence—such as that generated at second order by amplified scalar perturbations~\cite{Domenech:2021ztg,LISACosmologyWorkingGroup:2025vdz}—might offer much greater prospects for the detection of this kinematic boost than a scale invariant one. 

This speculation, however, is not valid beyond the Taylor-expanded relations~\eqref{boost}-~\eqref{MDQ-Omega}. In this work, we have demonstrated that the enhancement is absent when the exact boost transformation, eqs.~\eqref{Hcal_transform}, is taken into account. The failure of the Taylor expansion arises because higher-order terms in the small boost parameter $\beta$ involve increasingly higher derivatives of the spectral shape. As a result, truncating the expansion - typically done in the literature by retaining only the leading non-vanishing contribution to each multipole - is inadequate precisely in those scenarios where it seems to predict an enhanced boosted signal.

Fortunately, we can conclude on a more positive note. The relation~\eqref{boost} has the advantage of yielding a direct multipole expansion of the boosted signal, which can be conveniently used in proposed pipelines for analysing an anisotropic SGWB, where the detector response function is likewise expanded into multipoles~\cite{LISACosmologyWorkingGroup:2022kbp}. The lack of convergence in the Taylor-expanded relations~\eqref{MDQ-Omega} for strongly scale-dependent signals may raise concerns that existing pipelines are not applicable to such models. However, the exact relation~\eqref{Hcal_transform} remains a simple analytic expression for the SGWB in the boosted frame, which can be numerically expanded into multipoles in a straightforward and highly efficient manner. Once this expansion is performed, the standard analysis pipelines can be directly applied, without concern for spurious effects.

\begin{acknowledgments}
We thank Jacopo Fumagalli, Henri Inchauspé and Gianmassimo Tasinato for insightful discussions.
G.~M. acknowledges support from the Imperial College London Schr\"odinger Scholarship scheme. M.P. acknowledges support from Istituto Nazionale di Fisica Nucleare (INFN) through the Theoretical Astroparticle Physics (TAsP) project, and from the MIUR Progetti di Ricerca di Rilevante Interesse Nazionale (PRIN) Bando 2022 - grant 20228RMX4A.
\end{acknowledgments}

\bibliographystyle{apsrev}
\bibliography{paper-biblio}
	
\end{document}